# New compound fractional sliding mode control and super-twisting control of a MEMS gyroscope

Mehran Rahmani

*Abstract*— In this research we propose a new compound Fractional Order Sliding Mode Controller (FOSMC) and Super-Twisting Controller (FOSMC+STC) to control of a MEMS gyroscope. A new sliding mode surface has been defined to design the proposed new sliding mode controller. The main advantages of a FOSMC is its high tracking performance and robustness against external perturbation, but it is susceptible to chattering. By augmenting a STC with a FOSMC, the chattering phenomenon is eliminated, singularity problem is solved and systems robustness has significatnetly improved. Simulation results validate the effectiveness of the proposed control approach.

## I. INTRODUCTION

MICRO electro-mechanical system (MEMS) gyroscope are usually used as sensors for measuring angular velocity in stabilization applications with closed loop control. The performance of the MEMS gyroscope is impacted by mismatch in the frequency of oscillation between the two vibrating axes resulting from the effect of external disturbances and time varying model parameters [1]. Performance can be improved by designing a closed loop controller to negate the effects of external disturbances and time varying model parameters . In this paper we develop a novel the Fractional-Order Sliding Mode Controller (FOSMC) combined with a Super-Twisting Controller (STC) to control a MEMS gyroscope

Yang and Liu numerically considered the FOSMC for a new hyperchaotic structure [4]. Gao and Liao presented integral sliding mode control to enhance robustness of FOSMC [5]. Balochian used variable structure control for an individual polytopic system with fractional order operator. A specific feedback law is considered by proposing a sliding surface with fractional order operator [6]. Rabah et al. guaranteed asymptotic stability of fractional systems, by provided a novel technique of FOSMC [7]. Shah and Mehta described Thiran's delay estimation scheme in order to compensate the controller for actuator fractional delay. This considered the real time networked medium and packet loss situation [8]. Sun and Mah applied a fractional integral sliding mode control for tracking control of a linear motor in order to achieve high convergence precision,. Experimental results validated that the proposed control law has high tracking performance in comparison with conventional sliding mode control [9]. Wang et al. proposed a new fractional order nonsingular terminal sliding mode control. The proposed controller due to the fractional order nonsingular terminal sliding mode controller and fast terminal sliding mode controller, gaurantees fast convergence and high tracking performance [10]. Aghababa presented a new fractional hierarchical terminal sliding mode surface, where finite time convergence to the origin was demonstrated. A robust sliding mode switching control method was deisgned to guarantee the fractional Lyapunov stability [11].. Wang et al. implemented a novel sliding mode controller for an active vehicle suspension system to suppress the effect of external noise [12]. Based on aforementioned researches , FOSMC can be used as a strong control method in different systems [13-15], but its main drawback is creating chattering phenomenon STC overcomes disturbances and eliminated chattering phenomenon. Guruganesh et al. designed a control method for Micro Aerial vehicle using the second sliding mode control and super-twisting control [16]. Jeong et al. designed a robust super-twisting sliding mode control that guarantees high tracking trajectory of a robotic system. In order to satisfy the properties of a conventional sliding mode control, a super-twisting sliding mode surface is designed for obtaining the transient and steady-state time performances of the position of robotic manipulator [17]. Chuei et al. described a super-twisting observer based repetitive control, which overcome aperiodic disturbances [18]. Zargham and Mazinan applied super-twisting sliding mode control method to control of wind turbine system. Conventional sliding mode control is not guaranteed to maintain the closed loop performance against external pertubations. As a result of this drawback, a STC technique used for rapid response and high accuracy in chattering reduction [19]. Zhao et al., proposed a non-singular terminal sliding mode control based on STC method [20]. Lu and Xia addressed a new adaptive super-twisting algorithm for

Mehran Rahmani is with the Department of Mechanical Engineering, University of Wisconsin-Milwaukee, WI, CO 53206 USA (email: mrahmani@uwm.edu).

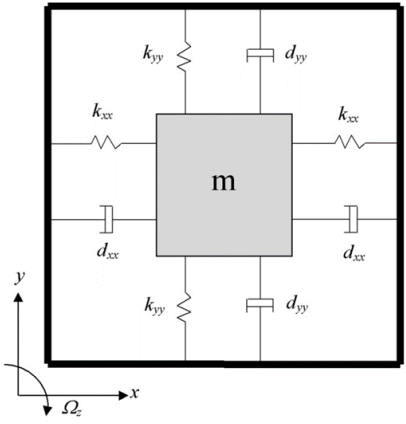

Fig. 1. Structure of a MEMS gyroscope.

control of rigid spacecraft. The applied controller is anti-singuarity and anti-chattering when encountered with external disturbances [21]. Evangelista et al. used modified STC to improve robustness of the system in the presence of external perturbations acting on a wind turbine shaft [22]. Becerra et al. proposed a STC which guarantees continuous control inputs and enhance robustness properties [23]. Salgado et al. introduced a discrete time super-twisting algorithm to solve the problems of control and state estimation [24]. As a result of considered studies, STC can be utilized as a strong tool in control systems. It includes some advantages such as improves robustness of control systems, removes controller singularity, and eliminates chattering phenomenon. In this research a novel FOSMC is proposed, which is robust against external perturbations and shows high tracking performance. The main drawbacks of FOSMC is creating chattering phenomenon. By using STC, it continuously calculates an error value and applies a correction value to the system. This process will eliminate chattering phenomenon and improve the robustness of the system.

In next section of the pape, dynamic modeling of a MEMS gyroscope is presented. In Section 3, the FOSMC is described. In Section 4, compound FOSMC+STC has been delineated. Section 5 presents simulation results. Finally the paper ends with the conclusion and the contributions of the rsearch work in the Setion 6.

## II. DYNAMIC OF MEMS GYROSCOPE

A z-axis MEMS gyroscope is shown in Figure 1. The conventional MEMS vibratory gyroscope consists of a proof mass (m) suspended by springs, where x and y are the coordinates of the proof mass with respect to the gyro frame in a Cartesian coordinate system, sensing mechanisms, and an electrostatics actuation for forcing an oscillatory motion and velocity of the proof mass and sensing the position. $\Omega x,y,z$ are the angular rate components along each axis of the gyro frame. The frame that the proof mass is mounted moves with a constant velocity and the gyroscope rotates at a slowly changing angular velocity $\Omega_z$. The centrifugal forces $m\Omega_z^2 x$ and $m\Omega_z^2 y$ are supposed to be negligible because of small displacements. The Coriolis force is generated in a direction perpendicular to the drive and rotational axes [25].

The dynamics of gyroscope according to assumptions presented above becomes as follows:

$$m\ddot{x} + d_{xx}^* \dot{x} + d_{xy}^* \dot{y} + k_{xx}^* x + k_{xy}^* y = u_x^* + 2m\Omega_z^* \dot{y} \quad (1)$$

$$m\ddot{y} + d_{xy}^* \dot{y} + d_{yy}^* \dot{y} + k_{xy}^* x + k_{yy}^* y = u_y^* - 2m\Omega_z^* \dot{x} \quad (2)$$

The origin for x, y coordinates is at the center of the proof mass without force employed. Fabrication imperfections is supposed to be helping basically to the asymmetric spring and damping terms, $k_{xy}^*$ and $d_{xy}^*$ respectively.

The x and y axes spring and damping terms $k_{xx}^*, k_{yy}^*, d_{xx}^*$ and $d_{yy}^*$ are often recognized, but may have small unknown variations from their nominal values [1, 25]. The mass of the proof mass m can be obtained exactly.

The positive direction of the control force is same as the x-y coordinate. $u_x^*$ and $u_y^*$ are the control forces in the x and y direction.

Dividing gyroscope dynamics (1) and (2) by the reference mass results in the following vector forms as:

$$\ddot{q}^* + \frac{D^*}{m}\dot{q}^* + \frac{K_a}{m}q^* = \frac{u^*}{m} - 2\Omega^* \dot{q}^* \quad (3)$$

where

$$q^* = \begin{bmatrix} x^* \\ y^* \end{bmatrix}, \quad u = \begin{bmatrix} u_x^* \\ u_y^* \end{bmatrix}, \quad \Omega^* = \begin{bmatrix} 0 & -\Omega_z^* \\ \Omega_z^* & 0 \end{bmatrix}$$

$$D^* = \begin{bmatrix} d_{xx}^* & d_{xy}^* \\ d_{xy}^* & d_{yy}^* \end{bmatrix}, \quad K_a = \begin{bmatrix} k_{xx}^* & k_{xy}^* \\ k_{xy}^* & k_{yy}^* \end{bmatrix}$$

The final form of the nondimensional equation of motion as follows:

$$\frac{\ddot{q}^*}{q_0} + \frac{D^*}{m\omega_0}\frac{\dot{q}^*}{q_0} + \frac{K_a}{m\omega_0^2}\frac{q^*}{q_0} = \frac{u^*}{m\omega_0^2 q_0} - 2\frac{\Omega^*}{\omega_0}\frac{\dot{q}^*}{q_0} \quad (4)$$

We determine a set of new parameters as follows:

$$q = \frac{q^*}{q_0}, \quad d_{xy} = \frac{d_{xy}^*}{m\omega_0}, \quad \Omega_z = \frac{\Omega_z^*}{\omega_0} \quad (5)$$

$$u = \frac{u_x^*}{m\omega_0^2 q_0}, \quad u_y = \frac{u_y^*}{m\omega_0^2 q_0} \quad (6)$$

$$\omega_x = \sqrt{\frac{k_{xx}}{m\omega_0^2}}, \quad \omega_y = \sqrt{\frac{k_{yy}}{m\omega_0^2}}, \quad \omega_{xy} = \frac{k_{xy}}{m\omega_0^2} \quad (7)$$

As a result, the nondimensional representation of (1) and (2) written as follows:

$$\ddot{q} + D\dot{q} + K_b q = u - 2\Omega \dot{q} \quad (8)$$

where

$$q = \begin{bmatrix} x \\ y \end{bmatrix}, \quad u = \begin{bmatrix} u_x \\ u_y \end{bmatrix}, \quad \Omega = \begin{bmatrix} 0 & -\Omega_z \\ \Omega_z & 0 \end{bmatrix}$$

$$D = \begin{bmatrix} d_{xx} & d_{xy} \\ d_{xy} & d_{yy} \end{bmatrix}, \quad K_b = \begin{bmatrix} \omega_x^2 & \omega_{xy} \\ \omega_{xy} & \omega_y^2 \end{bmatrix}$$

Clearly, Eq. (8) can be rearranged as:
$$\ddot{q} = -(D + 2\Omega)\dot{q} - K_b q + u + E \tag{9}$$

Where E is external disturbance. From Equation (9), the dynamic equations for a MEMS gyroscope becomes as:
$$\ddot{q} = -M\dot{q} - Nq + u + E \tag{10}$$

where $M = (D + 2\Omega)$ and $N = K_b$.

### III. NEW FRACTIONAL SLIDING MODE CONTROL

Selecting fractional sliding mode surface is the main part of FOSMC design. If a fractional sliding mode surface choose correctly, the best performance will be obtained. The new fractional sliding mode surface can be selected as follows:

$$s(t) = \dot{e}(t) + \alpha D^{\mu-1} e(t) + \beta D^{\mu-2} e(t) + \gamma \int_0^t e(\tau)^{\frac{r}{m}} d\tau \tag{11}$$

Where $r$, $m$, $\alpha$, $\beta$, and $\gamma$ are positive constant and $D$ is fractional order operator ($D = d/dt$, and $\mu > 2$). Fractional order operator type is Grunwald-Letnikov. The tracking error can be shown as:
$$e(t) = q - q_d \tag{12}$$

The derivative of fractional sliding mode surface is:
$$\dot{s}(t) = \ddot{e}(t) + \alpha D^{\mu} e(t) + \beta D^{\mu-1} e(t) + \gamma e(t)^{\frac{r}{m}} \tag{13}$$

$$= \ddot{q} - \ddot{q}_d + \alpha D^{\mu} e(t) + \beta D^{\mu-1} e(t) + \gamma e(t)^{\frac{r}{m}}$$

$$= -M\dot{q} - Nq + u + E - \ddot{q}_d + \alpha D^{\mu} e(t)$$

$$+ \beta D^{\mu-1} e(t) + \gamma e(t)^{\frac{r}{m}}$$

Equivalent control can be obtained by setting $\dot{s}(t) = 0$.

$$u_{eq}(t) = M\dot{q} + Nq - E + \ddot{q}_d - \alpha D^{\mu} e(t) \tag{14}$$

$$- \beta D^{\mu-1} e(t) - \gamma e(t)^{\frac{r}{m}}$$

The FOSMC can be shown as:
$$u_{FOSMC}(t) = u_{eq}(t) + u_s(t) = M\dot{q} + Nq - E + \ddot{q}_d \tag{15}$$

$$- \alpha D^{\mu} e(t) - \beta D^{\mu-1} e(t) - \gamma e(t)^{\frac{r}{m}} - K_s s$$

The equivalent control cannot compensate external perturbation and unmodel dynamic uncertainties. A reaching control law can be designed in order to remove those problems as $u_s(t)$, which can be defined as:
$$u_s(t) = -K_s s \tag{16}$$

where $K_s$ is positive constant.

Lyapunov theory can be taken into considerations as a strong tool for stability proving. Considering the following Lyapunov function candidate ($V$) which is continuous and nonnegative [26-28].

$$V = \frac{1}{2} s^T s \tag{17}$$

The time derivative of $V$ yields
$$\dot{V} = s^T \dot{s} = -M\dot{q} - Nq + u(t) + E - \ddot{q}_d \tag{18}$$

$$+ \alpha D^{\mu} e(t) + \beta D^{\mu-1} e(t) + \gamma e(t)^{\frac{r}{m}}$$

By substituting Eq. (15) into Eq. (18), generates
$$\dot{V} = s^T \dot{s} = s^T(-M\dot{q} - Nq + M\dot{q} + Nq - E + \ddot{q}_d \tag{19}$$

$$- \alpha D^{\mu} e(t) - \beta D^{\mu-1} e(t) - \gamma e(t)^{\frac{r}{m}}$$

$$- K_s s + E - \ddot{q}_d + \alpha D^{\mu} e(t)$$

$$+ \beta D^{\mu-1} e(t) + \gamma e(t)^{\frac{r}{m}})$$

Simplify Eq. (19) results in:
$$\dot{V} = s^T(-K_s s) \tag{20}$$

Therefore, Eq. (20) can be expressed as:
$$\dot{V} = -K_s s^2 \tag{21}$$

The Eq. (21) shows that $\dot{V} < 0$, which expressed that the proposed control law is stable.

### IV. NEW COMPOUND CONTROL SYSTEM

FOSMC is a strong tool which is able to enhance robustness of control system and improve tracking performance.
It's main drawbacks is creating chattering phenomenon. However, STC can be used in control systems as a powerful control approach, which its advantages can be enumerated as enhance robustness of the system, improve trajectory tracking, removing singularity problem, and eliminate chattering phenomenon.
By combining both FOSMC and STC, a new control method will be obtained which benefits both controller properties.
The compound control law can be defined as:
$$u(t) = u_{FOSMC}(t) + u_{STC}(t) \tag{22}$$

where $u_{STC}(t)$ is

$$u_{STC}(t) = -k_1 \left| e^{\frac{r}{m}} \right|^{\frac{1}{2}} sign(e^{\frac{r}{m}}) - k_2 \int_0^t sign(e^{\frac{r}{m}}) d\tau \tag{23}$$

where $k_1$, $k_2$, $r$, and $m$ are positive constants. The stability proving of the proposed control law can be arranged by substituting Eq. (22) into Eq. (18) as:
$$\dot{V} = s^T \dot{s} = s^T(-M\dot{q} - Nq + u_{FOSMC}(t) + u_{STC}(t) \tag{24}$$

$$+ E - \ddot{q}_d + \alpha D^{\mu} e(t) + \beta D^{\mu-1} e(t) + \gamma e(t)^{\frac{r}{m}})$$

Eq. (24) can be expressed as:

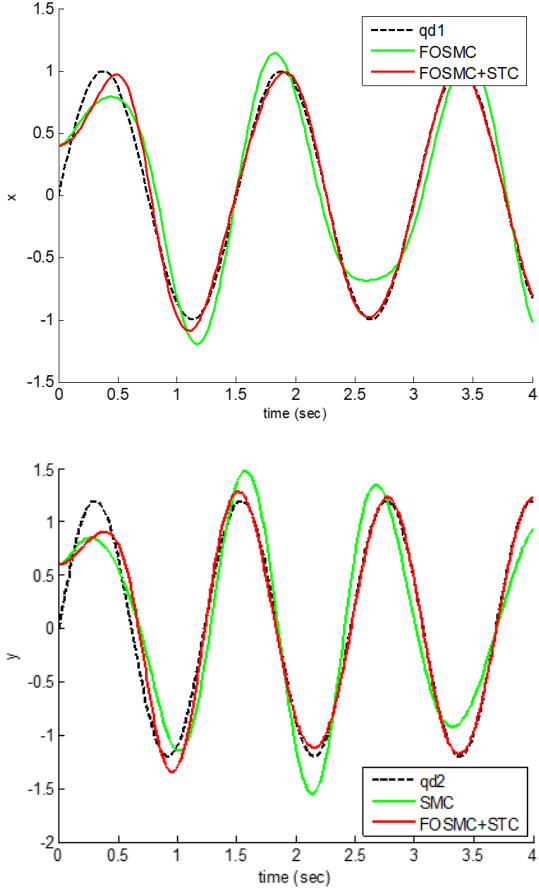

Fig. 2. Position tracking of x-axis and y-axis.

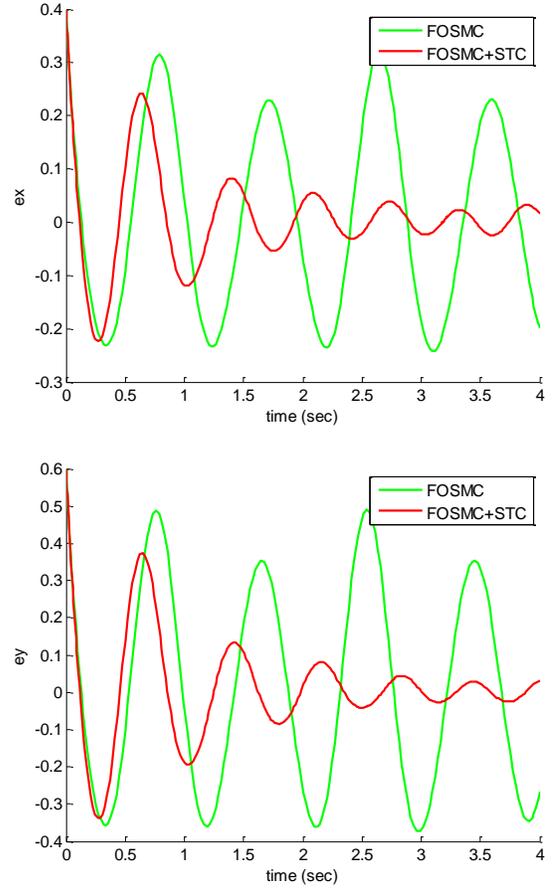

Fig. 3. Position tracking of x-axis and y-axis.

$$\dot{V} = s^T \dot{s} = s^T(-M\dot{q} - Nq + M\dot{q} + Nq \qquad (25)$$
$$- E + \ddot{q}_d - \alpha D^\mu e(t) - \beta D^{\mu-1} e(t)$$
$$- \gamma e(t)^{\frac{r}{m}} - K_s s - k_1 \left| e^{\frac{r}{m}} \right|^{\frac{1}{2}} sign(e^{\frac{r}{m}})$$
$$- k_2 \int_0^t sign(e^{\frac{r}{m}}) d\tau + E - \ddot{q}_d + \alpha D^\mu e(t)$$
$$+ \beta D^{\mu-1} e(t) + \gamma e(t)^{\frac{r}{m}})$$

Simplifying Eq. (25) generates.

$$\dot{V} = s^T(-K_s s - k_1 \left| e^{\frac{r}{m}} \right|^{\frac{1}{2}} sign(e^{\frac{r}{m}}) \qquad (26)$$
$$- k_2 \int_0^t sign(e^{\frac{r}{m}}) d\tau)$$

racking error will tend to zero ($e(t) \to 0$) when time goes to infinity ($t \to \infty$). Therefore, Eq. (26) can be written as:
$$\dot{V} = -K_s s^2 \qquad (27)$$

where $K_s$ is positive and $s_2 > 0$, the $\dot{V} < 0$ will be obtained

## V. SIMULATION RESULTS

Selection of proposed controller parameters (α, β, γ, Ks, μ, r, m, k1, and k2) is the most important part of the controller design procedure.

If parameters are chosen inappropriately, the proposed control method cannot guarantees the desired performance such as trajectory tracking, robustness, stability, and chattering elimination.

The controller parameters are chosen based on designer's experiences and trial-error process. Simulation results have shown that the parameters are selected appropriately.

Parameters of the fractional order sliding mode surface are selected as α=diag(40,40), β=diag(50,50), γ=diag(60,60), Ks=diag(10,10), μ=2.5, r=1.5 and m=1.25.

The STC parameters are chosen as k1=diag(20,20) and k2=diag(20,20) .The desired motion trajectory is determined by qd1=sin (4.17t), and qd2=1.2sin(5.11t).

The initial values of the system are selected as $q_1(0) = 0.5, q_2(0) = 0.5, \dot{q}_1(0) = 0 \text{ and } \dot{q}_2(0) = 0$

The parameters of the MEMS gyroscope are selected as:

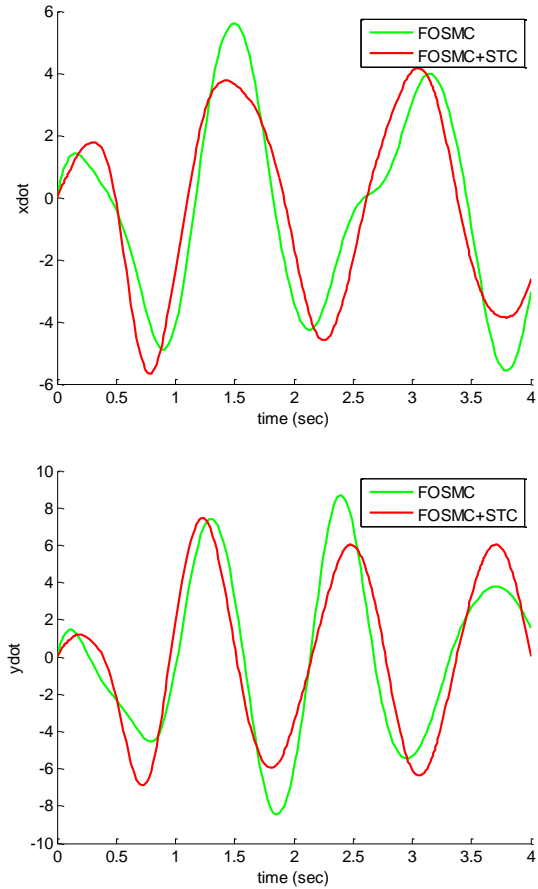

Fig. 5. Velocity of x-axis and y-axis.

$m = 1.8 \times 10^{-7} kg \qquad k_{xy} = 12.779 N/m$

$k_{xx} = 63.955 N/m \qquad d_{xx} = 1.8 \times 10^{-6} Ns/m$

$k_{yy} 95.92 N/m \qquad d_{yy} 1.8 \times 10^{-6} Ns/m$

$d_{xy} = 3.6 \times 10^{-7} Ns/m$

The conventional natural frequency of each axis of a MEMS gyroscope is in the KHZ range, so, choose the ω0 as 1KHZ. It is suitable to choose 1μm as the reference length q0 when the displacement rang of the MEMS gyroscope in each axis is sub-micrometer level. The unknown angular velocity is assumed Ωz=100 rad/s. Therefore, the non- dimensional values of the MEMS gyroscope parameters are chosen as:

$\omega_x^2 = 355.3, \omega_y^2 = 532.9, \omega_{xy} = 70.99, d_{xx} = 0.01, d_{yy} = 0.01,$

$d_{xy} = 0.002, \Omega_z = 0.1$

Figure 2 shows position tracking of x-axis and y-axis under FOSMC and FOSMC+STC. It can be seen clearly that tracking performance under proposed controllers is consistent with desired tracking for the MEMS gyroscope. Figure 3 illustrates tracking error of x-axis and y-axis under FOSMC and proposed control. FOSMC creates chattering phenomenon, which by using STC, chattering phenomenon reduced. In addition, FOSMC+STC has lower maximum overshoot and undershoot in comparison with FOSMC. Figure 4 shows velocity of x-axis and y-axis under FOSMC and proposed control law.

## VI. CONCLUSION

In this study, we proposed a novel FOSMC+STC law for control of a MEMS gyroscope. A new FOSMC is applied for control of x-axis and y-axis of a MEMS gyroscope. It has high tracking performance, but its main drawbacks was creating chattering phenomenon. In order to solve this problem, a STC is proposed in parallel with FOSMC, which continuously claculates an error value and applies a correction value. Simulated resuts demonstate that the developed STC significantely reduce the chattering phenomenon. In addition, by using STC, maximum overshoot and undershoot reduce and trajectory tracking performance improved. Simulation results thus validated the effectiveness of the proposed control strategy.